\documentclass[twocolumn,english,showpacs,floatfix,amsmath,amssymb,prl,reprint]{revtex4-1}
\usepackage[T1]{fontenc}
\usepackage[latin9]{inputenc}
\usepackage{babel}

\usepackage{graphicx}
\usepackage{amssymb}
\usepackage{esint}
\renewcommand{\vr}{{\bf r}}

\usepackage[usenames]{color}
\usepackage{soul}
\setulcolor{red}
\setstcolor{red}
\sethlcolor{yellow}



\begin{document}

\title{Enhanced zero-bias Majorana peak in disordered multi-subband quantum wires}

\author{Falko Pientka} 
\affiliation{\mbox{Dahlem Center for Complex Quantum Systems and Fachbereich Physik, Freie Universit\"at Berlin, 14195 Berlin, Germany}}

\author{Graham Kells}
\affiliation{\mbox{Dahlem Center for Complex Quantum Systems and Fachbereich Physik, Freie Universit\"at Berlin, 14195 Berlin, Germany}}

\author{Alessandro Romito}
\affiliation{\mbox{Dahlem Center for Complex Quantum Systems and Fachbereich Physik, Freie Universit\"at Berlin, 14195 Berlin, Germany}}

\author{Piet W.\ Brouwer} 
\affiliation{\mbox{Dahlem Center for Complex Quantum Systems and Fachbereich Physik, Freie Universit\"at Berlin, 14195 Berlin, Germany}}

\author{Felix von Oppen}
\affiliation{\mbox{Dahlem Center for Complex Quantum Systems and Fachbereich Physik, Freie Universit\"at Berlin, 14195 Berlin, Germany}}

\date{\today}
\begin{abstract}
A recent experiment [Mourik {\em et al.}, Science  {\bf 336}, 1003 (2012)] on InSb quantum wires provides possible evidence for the realization of a topological superconducting phase and the formation of Majorana bound states. Motivated by this experiment, we consider the signature of Majorana bound states in the differential tunneling conductance of multi-subband wires. We show that the weight of the Majorana-induced zero-bias peak is strongly enhanced by mixing of subbands, when disorder is added to the end of the quantum wire. We also consider how the topological phase transition is reflected in the gap structure of the current-voltage characteristic. 
\end{abstract}
\maketitle

{\em Introduction.---}A recent experiment \cite{kouwenhoven12} reports the realization of proximity-induced topological superconductivity \cite{potter10,fu08,sau10} and the formation of Majorana bound states in InSb quantum wires. Following theoretical suggestions \cite{lutchyn10,oreg10}, superconducting order is induced in an InSb quantum wire by proximity to a Nb lead attached alongside the wire. At the other end, the quantum wire is contacted to a normal lead via a gate-induced tunnel junction. Evidence for the formation of Majorana bound states is found through measurements of the differential conductance, which exhibits a zero-bias peak when a magnetic field is applied in certain directions. Similar results were also obtained for normal-metal--superconductor structures based on InAs quantum wires \cite{das}.

At zero temperature and in single-subband quantum wires, the Majorana-induced zero-bias peak is predicted to have a height of $2e^2/h$ \cite{flensberg10,Law2009}.
At finite temperature, the zero-bias peak broadens with its weight fixed, so that the peak height is no longer expected to reach $2e^2/h$. In this paper, we consider the current-voltage characteristic of multi-subband wires --- a situation which is presumably relevant to the experiment of Ref.\ \cite{kouwenhoven12} ---, with a particular emphasis on the dependence of the zero-bias peak on subband mixing by disorder. 
We show that, remarkably, the weight of the Majorana-induced zero-bias peak is typically enhanced as the tunnel junction becomes more disordered. The basic idea is that disorder couples the topological channel, which itself is only weakly transmitted through the barrier, with the other non-topological subbands which have higher transmission coefficients. This coupling broadens the conductance peak and hence, in the presence of a finite temperature, enhances the zero-bias conductance \cite{foot}. The intentional inclusion of disorder in or near the barrier, either during the fabrication process of the InSb nanowires used in the experiment, or after fabrication of the device, could thus lead to an additional, strong signature of the Majorana end state. A similar effect is expected if the tunnel barrier is replaced by a point contact \cite{wimmer11}, provided the point contact is non-adiabatic.

The interest in Majorana bound states in low-dimensional condensed matter systems \cite{review12,kouwenhoven12,Kitaev2001,read00,goldhaber,deng,rokhinson,das} is driven by their remarkable properties: They are their own antiparticle, have zero energy, and obey non-Abelian exchange statistics \cite{moore91,ivanov} upon adiabatic permutation of their positions. The latter two properties make Majorana bound states potentially useful for topological quantum computation \cite{kitaev03}. 

\begin{figure}[b]
\includegraphics[width=.48\linewidth, keepaspectratio=true]{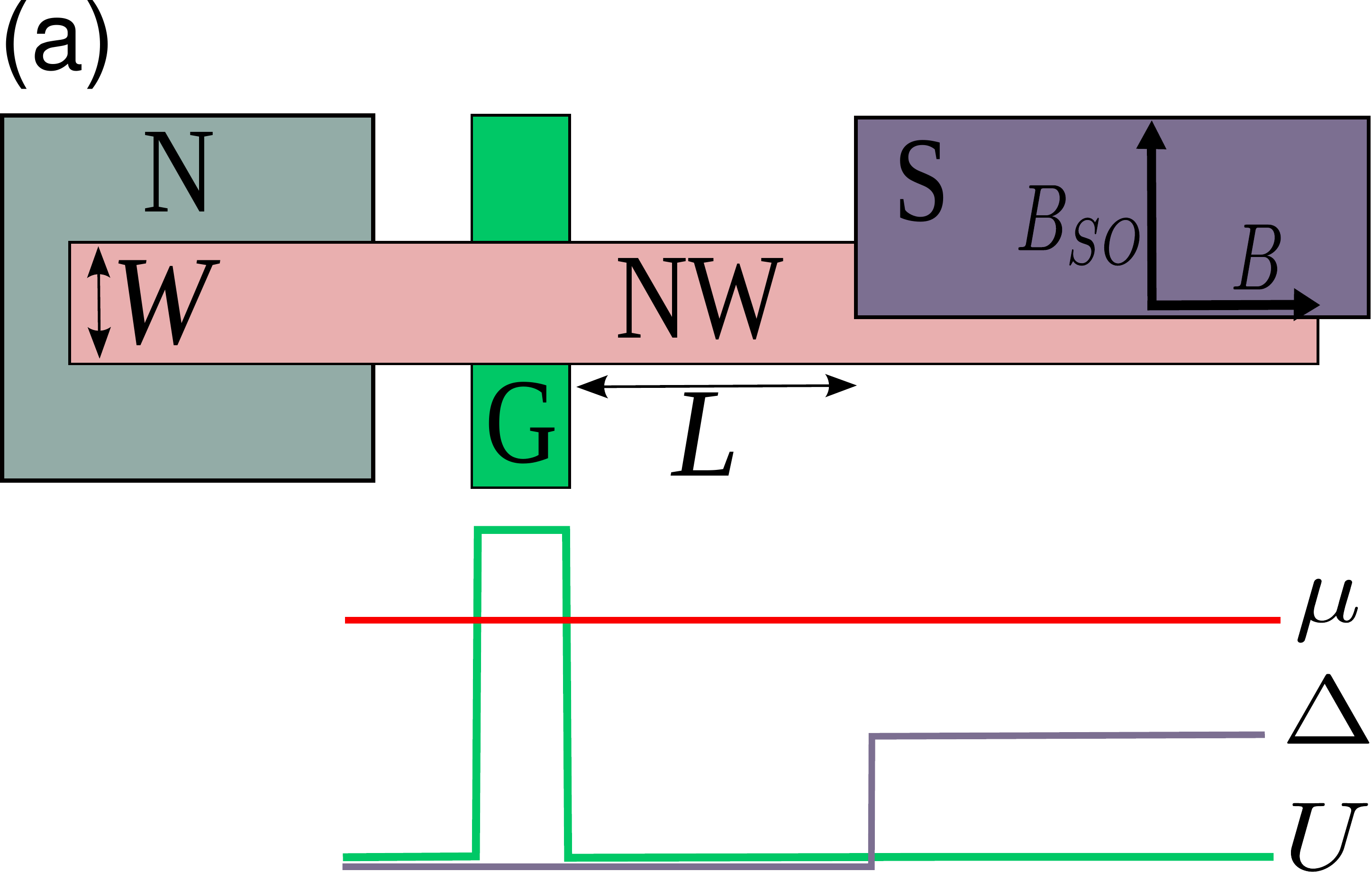} 
\includegraphics[width=.48\linewidth, keepaspectratio=true]{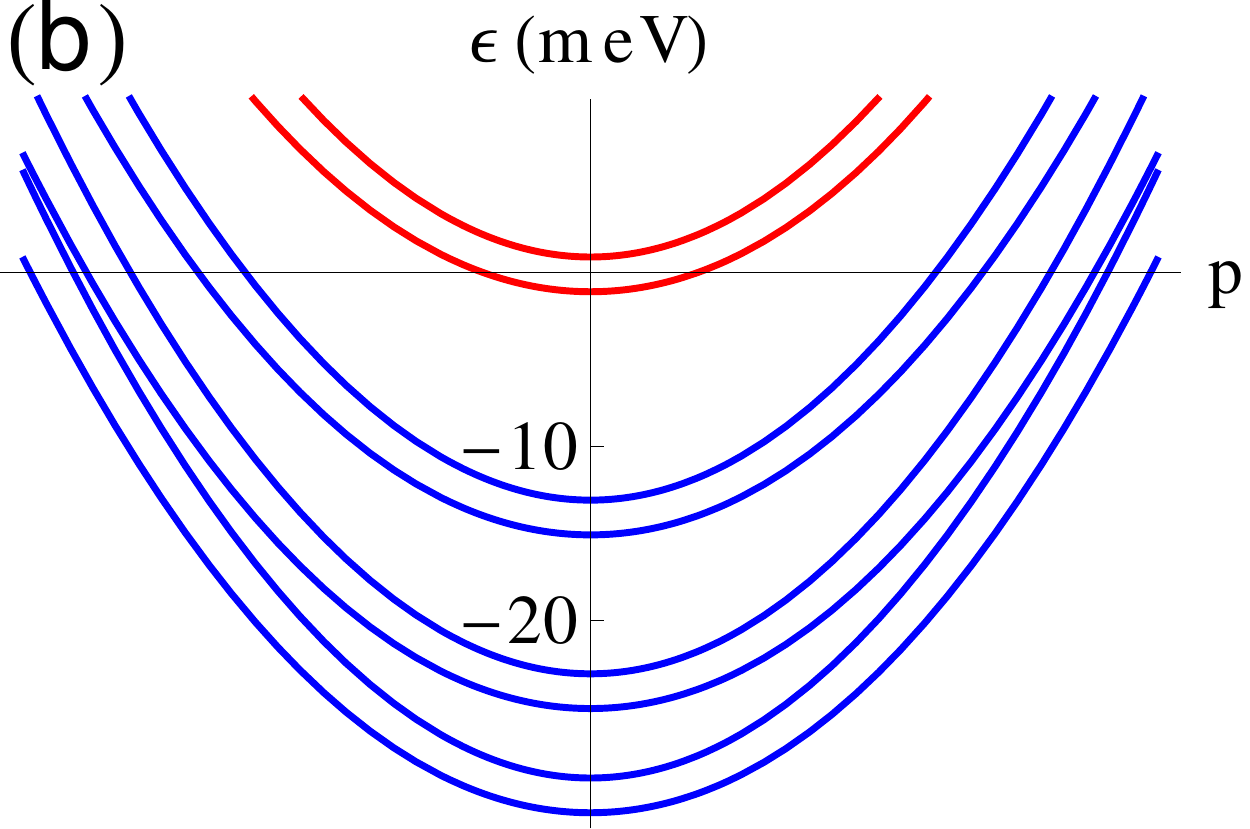} 
\caption{(Color online)
(a) Setup of multi-subband quantum nanowire (NW) with gate-induced tunnel barrier (G) and proximity coupled $s$-wave superconductor (S). As in Ref.\ \cite{kouwenhoven12} we consider the conductance between the normal lead (N) and the superconductor. Subband mixing is induced through disorder in the short segment of length $L$ between the tunnel barrier and superconductor. (b) Normal-state dispersion in the absence of disorder for four subbands with $B=1$ meV and $m\alpha^2/2=50$ $\mu$eV.
\label{fig:setup}}
\end{figure}

{\em Model system}.---We consider a geometry close to that of the experiment in Ref.~\cite{kouwenhoven12} shown schematically in Fig.~\ref{fig:setup}a. It consists of a two-dimensional multi-subband semiconducting wire with spin-orbit velocity $\alpha$, chemical potential $\mu$, and width $W$. At one end, the semiconductor is coupled laterally to a superconducting lead. At the other end, it is contacted to a normal metal via a tunnel barrier defined by the gate potential $U$. The system is placed in a magnetic field parallel to the wire direction with Zeeman energy $B$. Taking the $x$ direction to be along the wire, the system is then described by the Bogoliubov--de Gennes Hamiltonian \cite{oreg10,lutchyn10,Lutchyn2011,review12} 
\begin{align}
  {\cal H} = & \left( \frac{{\bf p}^2}{2m} + \alpha p_x \sigma_y - \alpha_y p_y\sigma_x + U(x) +V_{\rm dis}({\bf r}) - \mu \right)\tau_z\nonumber\\
  &\quad - B \sigma_x + \Delta(x) \tau_x ,
  \label{BdG}
\end{align}
where the Pauli matrices $\sigma$ and $\tau$ operate on the spin and particle-hole degrees of freedom, respectively. The parameter $\alpha_y$ is included for future reference and equals $\alpha$ for the case of Rashba spin-orbit coupling. The lateral contact to the superconductor covers the region $x > 0$, so that we set $\Delta(x) = \Delta$ for $x > 0$ and $\Delta(x)=0$, otherwise, where $\Delta$ is the proximity-induced gap for $B=0$. The disorder potential $V_{\rm dis}({\bf r})$ is nonzero in the region $-L < x < 0$ between the gate-defined tunnel barrier and the superconducting contact only. In this region, we choose a Gaussian random potential with $\langle V_{\rm dis}(\vr) \rangle = 0$ and
\begin{equation}
  \langle V_{\rm dis}(\vr) V_{\rm dis}(\vr') \rangle =
 \frac{v_F^2}{k_Fl_{\rm 2d}} \delta(\vr-\vr'),
\end{equation}
where $v_F=\sqrt{2\mu/m}$ and $l_{\rm 2d}$ are the Fermi velocity and mean free path. 
It is important to note that the mechanism discussed here is different from reflectionless tunneling~\cite{reflectionless} induced by disorder on the normal side of $NS$ tunnel junctions.

We numerically calculate the normal and Andreev reflection matrices $r_{\rm ee}(\varepsilon)$ and $r_{\rm he}(\varepsilon)$ for the Hamiltonian (\ref{BdG}), using the technique described in Ref.~\cite{brouwer11a}. The differential conductance $G(V)$ is then evaluated according to \cite{Blonder1982}
\begin{align*}
 G(V)=\frac{e^2}{h} \mbox{tr}\, [1+ r_{\rm he}(eV) r_{\rm he}(eV)^{\dagger}-
  r_{\rm ee}(eV) r_{\rm ee}(eV)^{\dagger}],
\end{align*}
where the trace is in spin and channel space. In a wire of width $W$ lateral momenta are quantized as $p_{y,n}=\hbar n\pi/W$ with $n=1,2,\ldots$. In our numerical calculations we use an effective mass $m=0.015m_e$, $m_e$ being the bare electron mass, proximity induced gap $\Delta=250$ $\mu$eV, spin-orbit energy $m\alpha^2/2=50$ $\mu$eV, and width $W=110$ nm.~\cite{foot2}.
This choice corresponds to the parameters of the InSb quantum wires used in Ref.~\cite{kouwenhoven12}. The chemical potential in the nanowire is chosen as $\mu=32.1$ meV, corresponding to $N=4$ occupied channels (cf.~Fig.~\ref{fig:setup}b). The subbands in the nanowire are therefore separated by several meV and for Zeeman energies less than 1.5 meV (corresponding to  $B < 1$ T for InSb) only the highest channel (subband index $n=N=4$) can be in the topological phase.

\begin{figure}[t]
\includegraphics[width=.98\linewidth, keepaspectratio=true]{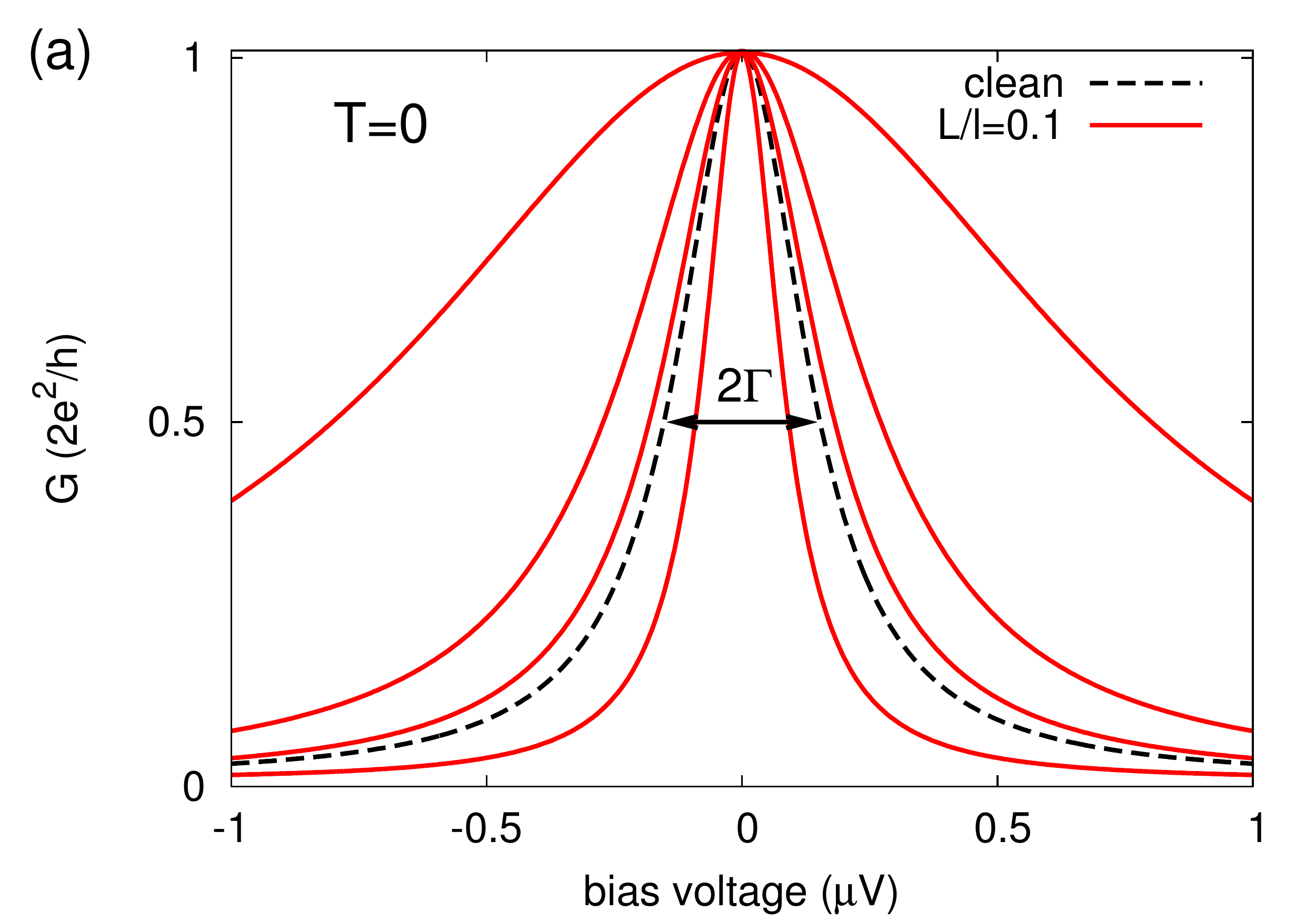}\\
\includegraphics[width=.98\linewidth, keepaspectratio=true]{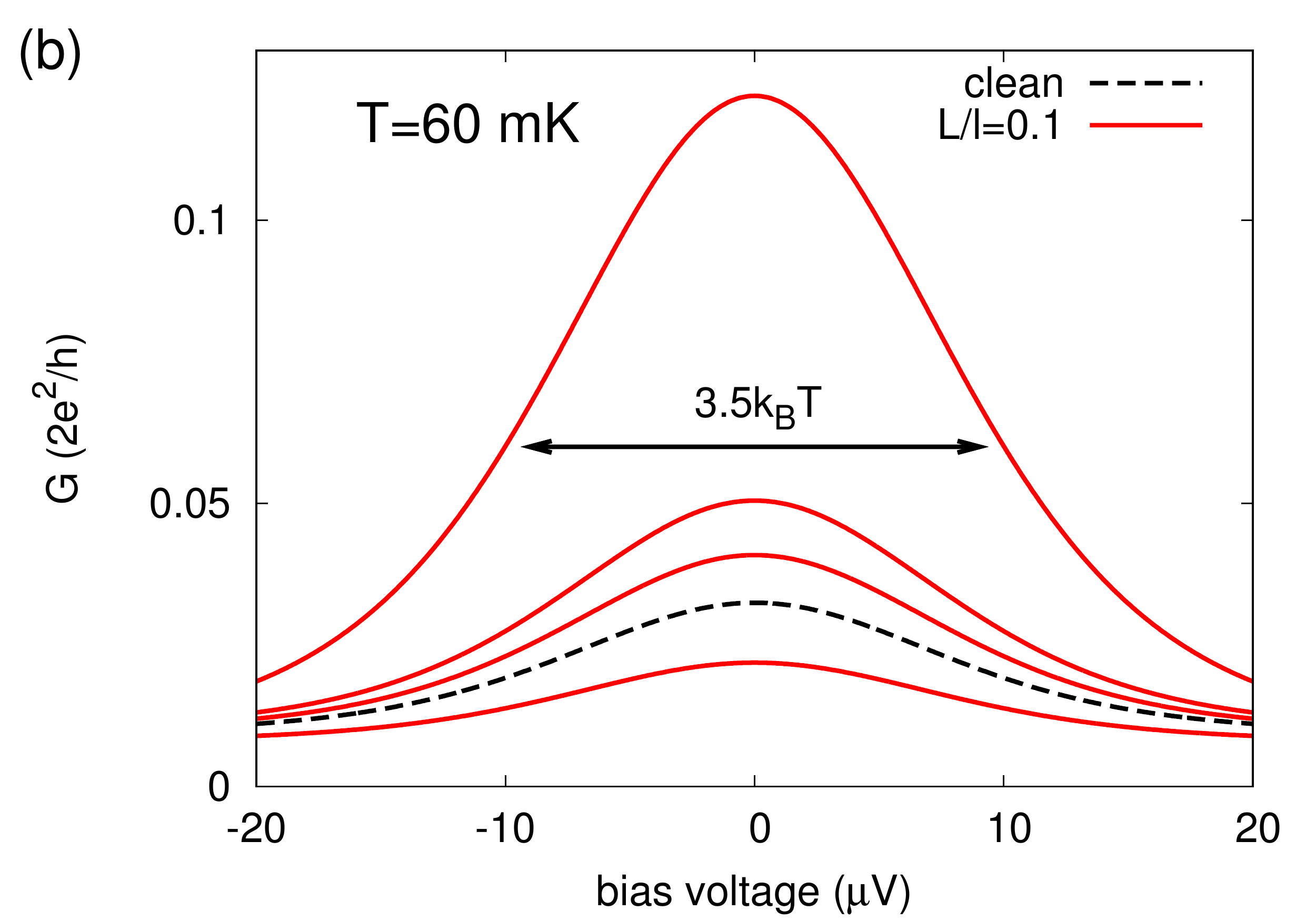} 
\caption{(Color online)
(a) Zero-bias conductance peak at zero temperature in a quantum wire with $B=0.5$ meV and $N=4$ transverse subbands, one of which is in the topological phase with barrier transmission $T_4=0.01$. The three nontopological subbands have transmissions 20$T_4$, 10$T_4$, and 4$T_4$.
The red curves show the conductance for four different disorder configurations with $l=10L$. The black dashed line shows the peak shape for the clean wire.
(b) Same as in (a), but for a temperature $T = 60$ mK, larger than the zero-temperature peak width.
\label{fig:temp}}
\end{figure}

{\em Clean multi-subband quantum wires.---}We first consider a clean multi-subband wire with $\alpha_y = 0$. To a good approximation, a gate-induced tunnel barrier exposes the electrons to a potential which depends only on the coordinate $x$ along the wire. Consequently, the tunnel barrier does not mix the transverse subbands (channels) of the quantum wire and the subbands can effectively be considered as independent. Each subband is characterized by a Fermi velocity $v_{F,n}=\frac{1}{m}(2m\mu-p_{y,n}^2)^{1/2}$, an excitation gap $\Delta_n$, and a transmission coefficient $T_n$ of the gate-induced tunnel barrier. 
Since the highest occupied subband $n=N$ determines whether the wire is in the topological phase, we refer to this subband as the ``topological subband.'' A nontrivial topological phase exists if $B^2 > B_N^2 = \Delta^2 + (\mu - p_{y,N}^2/2m)^2$ \cite{lutchyn10,oreg10}. At the topological phase transition, {\em i.e.}, for $B = B_N$, the topological gap $\Delta_N$ vanishes, whereas the excitation gaps for the other subbands remain finite. 

The topological phase is characterized by a zero-bias conductance peak
\begin{equation}
  G(V) = \frac{2 e^2}{h} \frac{\Gamma^2}{\Gamma^2 + (e V)^2},
\end{equation}
with width $\Gamma =\beta_0 \Delta_N T_N$ if $L\ll\xi=\hbar v_{F,N}/\Delta_N$. The numerical constant $\beta_0$ takes the value $\beta_0\approx 0.375$ for the range of parameters we investigated ($T_N\ll1$, $\Delta$ between 0 and 60 $\mu$eV). This width may be very small, since the transmission coefficient $T_N$ of the topological subband is typically much smaller than the transmission coefficients of the other channels. (For the $N$th subband to be topological, it is important that its band bottom be close to the chemical potential.) At finite temperature, the conductance peak is thermally broadened,
\begin{equation}
  G(V,T) = \int_{-\infty}^{\infty}d\varepsilon G(\varepsilon,0)\frac{d f}{d\varepsilon}(eV-\varepsilon,T),
\end{equation}
where $f (\epsilon,T)$ is the Fermi distribution function at temperature $T$. Thermal broadening preserves the weight of the zero-bias peak. For $k_{\rm B} T \gg \Gamma$, the peak width is of order $k_{\rm B} T$, whereas the height $ (2 e^2/h) (\pi \Gamma/4k_{\rm B} T)$ is inversely proportional to temperature. Both regimes are illustrated in Fig.~\ref{fig:temp}.

\begin{figure}[t]
\includegraphics[width=.98\linewidth, keepaspectratio=true]{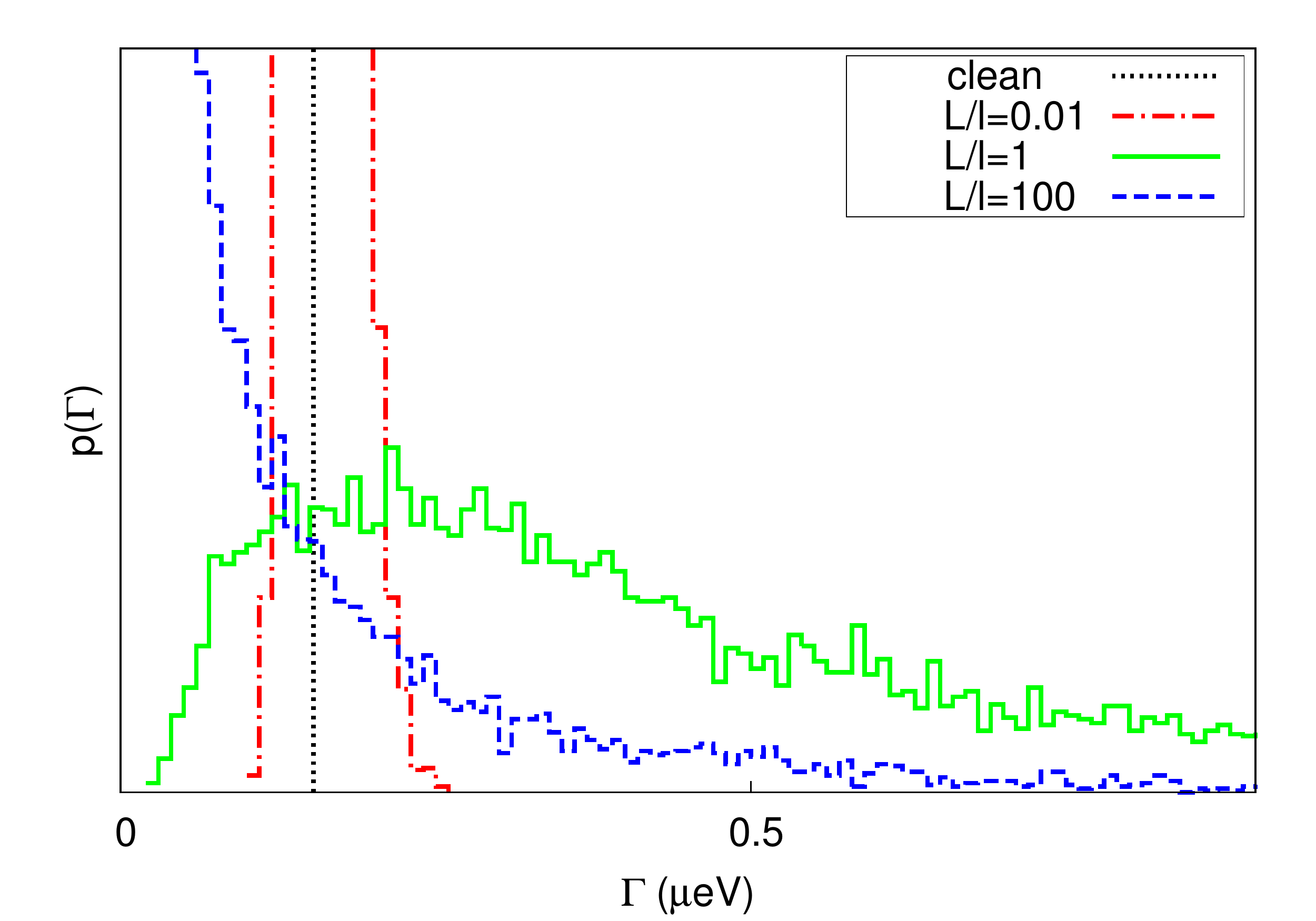}
\caption{(Color online)
Probability distribution of the zero-bias peak width $\Gamma$ in the presence of disorder in the nanowire segment $-L<x<0$ between the superconducting part and the barrier for a multi-channel wire with the same choice of parameters as in Fig.~\ref{fig:temp}. 
With increasing disorder, $\Gamma$ increases on average (red and green curve) due to subband mixing.
For $L\gg l$ Anderson localization reduces the overall transparency of the junction, causing $\Gamma$ to decrease again in the case of very strong disorder (blue curve).
\label{fig:hist}}
\end{figure}

\begin{figure}[t]
\includegraphics[width=.98\linewidth, keepaspectratio=true]{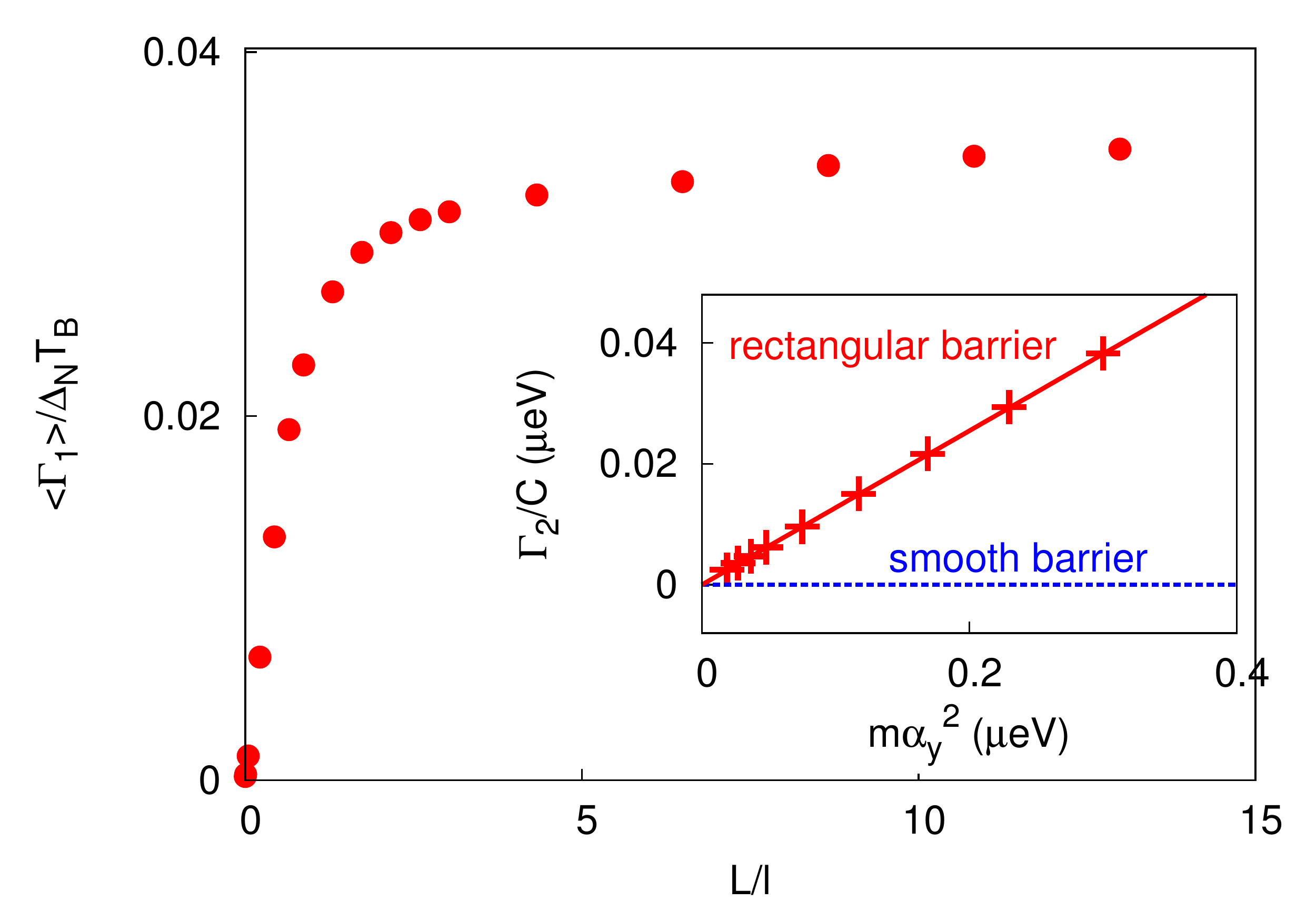} 
\caption{(Color online)
Ensemble average of the contribution $\Gamma_1$ from disorder-induced subband mixing to the width $\Gamma$ of the zero-bias peak as a function of disorder strength in the segment $-L<x<0$. The peak width is normalized by the normal state conductance $G_B\equiv (2e^2/h)NT_B$ to focus on the effects of subband mixing and to eliminate changes in the overall transparency by Anderson localization.
Inset: Contribution $\Gamma_2$ to the peak width from lateral spin-orbit coupling for a rectangular barrier (red crosses) and a Gaussian barrier (blue dashed line).
In both figures the parameters of the barrier potential have been chosen such that only $T_1$ differs appreciably from zero.
\label{fig:dis}}
\end{figure}

{\em Effect of disorder.---}If the disorder is limited to the segment of the semiconductor wire that is not in contact with the superconductor, i.e., to $-L < x < 0$ (cf.~Fig.~\ref{fig:setup}), it has no effect on the existence of the topological phase \cite{brouwer11b}. However, impurity scattering in the ``normal'' part of the wire has profound consequences for the weight of the zero bias peak associated with the existence of the topological phase. The underlying reason is the large disparity in the transparencies of the different subbands, with the topological subband having the smallest transparency $T_N$. Mixing of subbands by impurity scattering allows for the coupling of the topological subband to the normal lead via the lower subbands with higher transparency. 

The effect is illustrated in Fig.~\ref{fig:temp}, where we show the shape of the zero-bias peak for various disorder configurations, such that the distance $L$ between gate-induced tunnel barrier and the superconducting contact equals one tenth of the characteristic scattering length 
\begin{align}
l=l_{\rm 2d} v_{F,1}v_{F,N}/v_F^2. 
\end{align}
A systematic dependence on disorder strength can be seen in Figs.\ \ref{fig:hist} and \ref{fig:dis}. 
Figure \ref{fig:hist} shows the probability distribution of the zero-temperature peak width for different values of the ratio $L/l$. We conclude that already a moderate amount of disorder causes subband mixing and an increase in the peak width. At very strong disorder, $L \gg l$, Anderson localization suppresses the overall coupling to the normal lead, leading to a decrease of the weight of the zero-bias peak. This effect is not related to subband mixing and can be removed by normalizing the peak weight to the normal-state conductance $G_B\equiv(2e^2/h) NT_B$ of the device; see Fig.~\ref{fig:dis}. The average peak width from disorder-induced subband mixing saturates for $L/l\gg 1$ to $\left<\Gamma_1\right>=\beta_1 T_B\Delta_N $, where $\beta_1$ is a numerical factor of the order of $0.1$, the exact value depending on the barrier transparencies for different subbands and spin mixing due to the magnetic field.

{\em Other causes of subband mixing.}---The lateral spin-orbit term proportional to $\alpha_y$ in Eq.~(\ref{BdG}) may be an additional source of subband mixing. For small $\alpha_y$ its contribution to the width of the zero-bias peak is 
\begin{align}
 \Gamma_2=\beta_2 Cm\alpha_y^2,\qquad C= \frac{\Delta_NT_B}{W^2k_F^3v_{F,N}},
\end{align}
proportional to $\alpha_y^2$, with a numerical prefactor $\beta_2$ that depends on the precise shape of the barrier.
In the inset of Fig.~\ref{fig:dis} we show its effect on the conductance of a clean wire for a long and low tunnel barrier, so that only the lowest subband $n=1$ has an appreciable transmission. For a rectangular barrier, the subband mixing caused by lateral spin-orbit coupling is maximal, but still weak in comparison to the maximal subband mixing obtained from disorder, since for $B=0.5$ meV we obtain $C\approx 10^{-4}$. For a smooth barrier, which is the experimentally relevant limit, the numerical prefactor $\beta_2$ becomes vanishingly small and lateral spin-orbit coupling does not give any appreciable subband mixing.
Subbands may also be mixed by a gate-defined barrier that is not perpendicular to the direction of the wire. The mixing effect is maximal if the barrier is rectangular, and effectively absent for smooth barriers.

\begin{figure}[t]
\includegraphics[width=.98\linewidth, keepaspectratio=true]{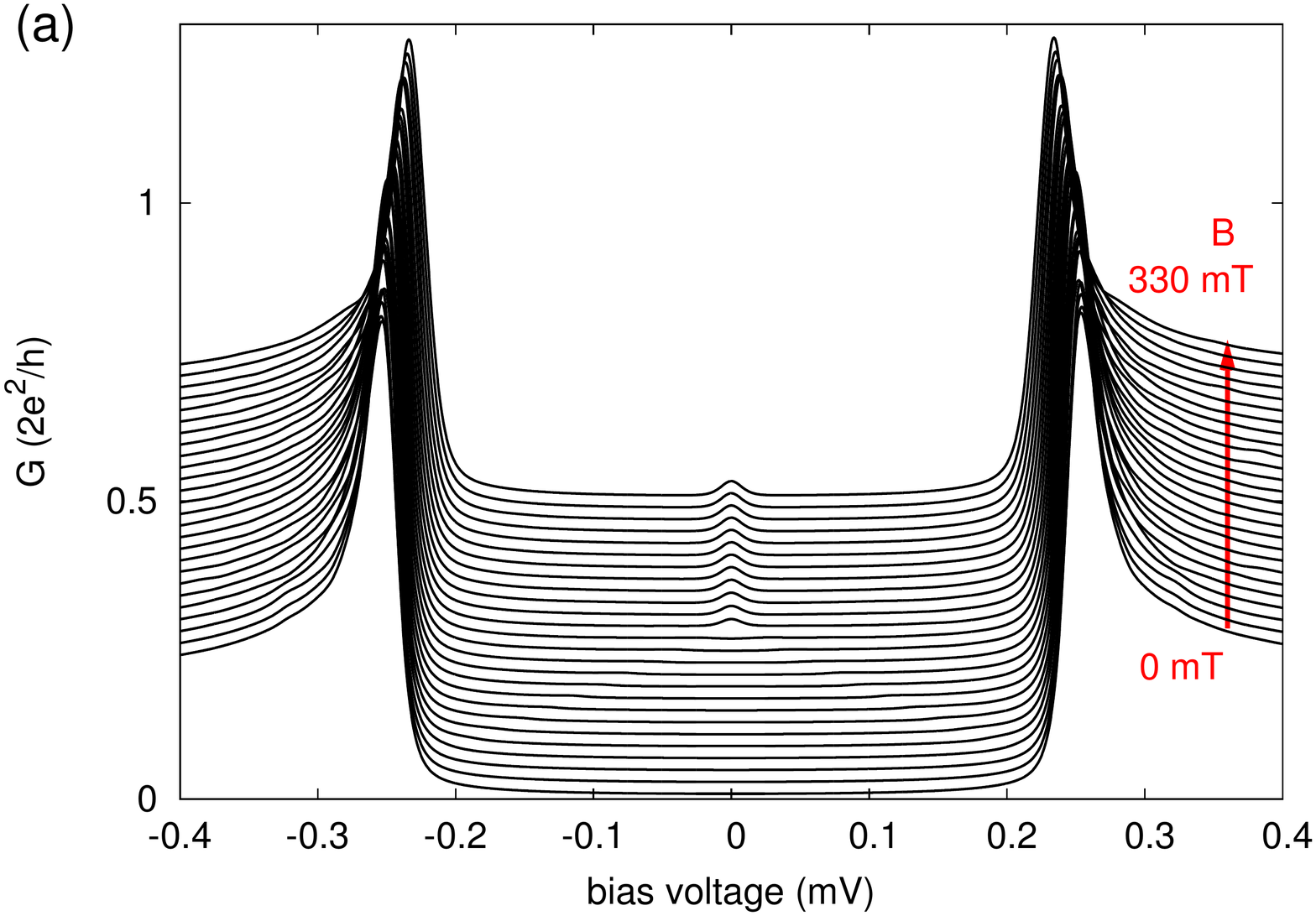} 
\includegraphics[width=.98\linewidth, keepaspectratio=true]{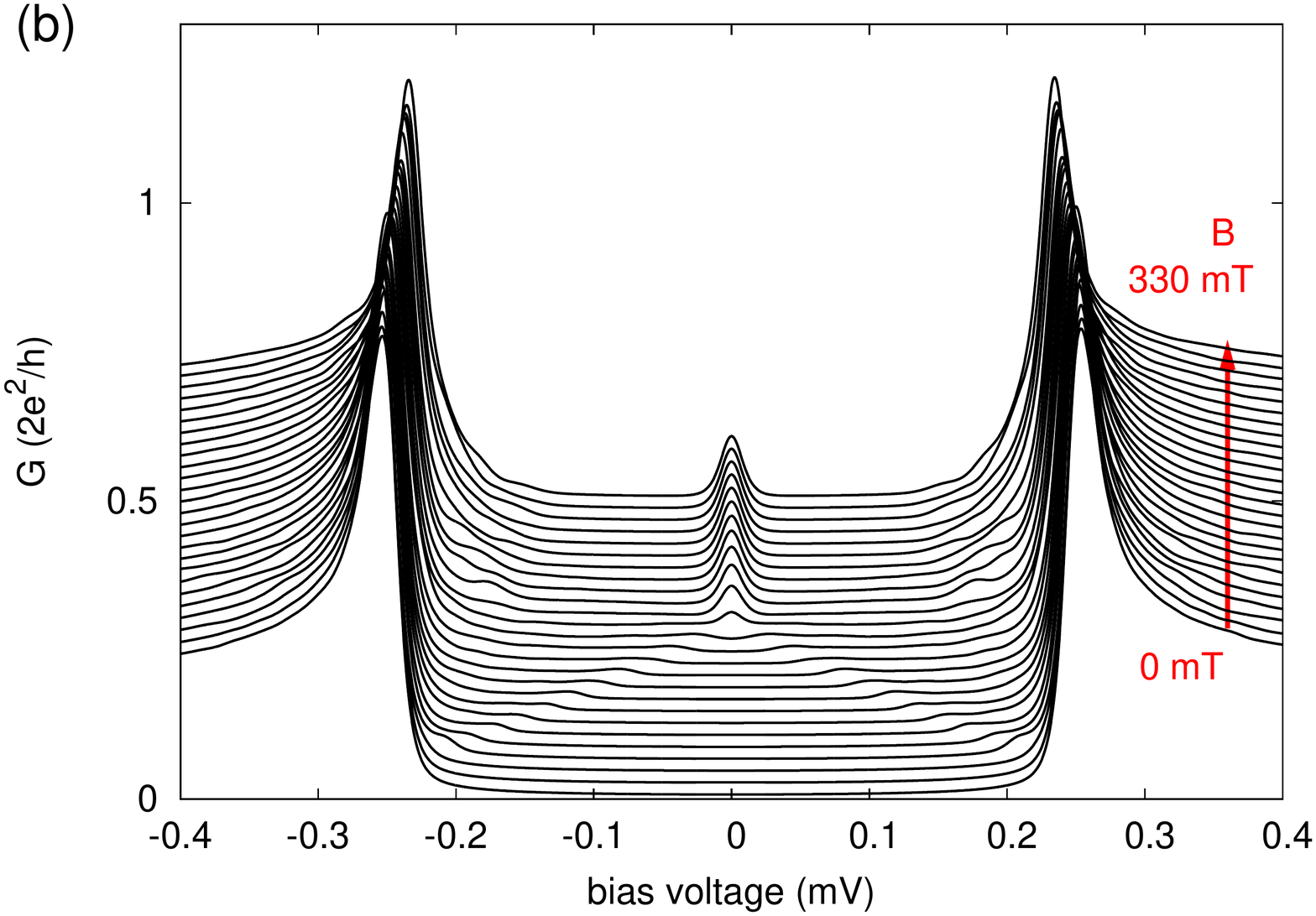} 
\caption{(Color online)
(a) Differential conductance vs bias voltage in a clean multichannel nanowire for increasing $B$ from 0 to 0.5 meV (750 mT in InSb) in steps of 0.02 meV with the realistic parameters \cite{kouwenhoven12} $\alpha_y=0$, $T=60$ mK ($k_{\rm B}T=5$ meV), $L=10$ nm, and $T_N=0.01$. 
The $B>0$ traces are offset vertically for clarity. The formation of a Majorana fermion is reflected in the emergence of a zero-bias peak. The corresponding closing of the topological gap is hardly discernible due to the low transparency of the topological channel. For $B=0$ there are coherence peaks at the proximity induced gap $\Delta=0.25$ meV. For larger Zeeman fields the bulk gap of the lower channels is decreased consistently with expectations.
(b) Same as (a) but with weak disorder in the region $-L<x<0$ adjacent to the barrier. All traces are calculated for the same disorder configuration with a scattering length $l=10L$. The zero-bias peak and the signature of the topological gap closing are considerably enhanced.
\label{fig:gaps}}
\end{figure}

{\em Current-voltage characteristic and topological gap.---}Unlike in single-channel models for spinless $p$-wave superconductors, multi-subband models are characterized by the coexistence of multiple superconducting gaps in different sections of the Fermi surface. Specifically, the proximity-induced gaps in the lower subbands are  only weakly affected by the applied magnetic field. In contrast, the highest occupied subband should have a gap closing when it enters into the topological superconducting phase at the critical magnetic field. Thus, it is interesting to investigate to which degree the differential conductance contains signatures of the gap closing at the topological phase transition and how disorder near the barrier affects these signatures.

In Fig.\ \ref{fig:gaps}a, we show the differential conductance versus bias voltage for a clean multichannel quantum wire at $T=60$ mK. At the critical Zeeman field of the topmost channel $B_c=0.27$ meV a peak appears at zero bias voltage. Since the topological channel is only weakly transmitted through the barrier, its contribution to the conductance is weaker than that of the other channels. 
In conjunction with density of states effects \cite{stanescu12}, this explains the very weak signature of the topological gap closing in transport in Fig.\ \ref{fig:gaps}a, consistent with the absence of the topological gap in the experimental measurements of Ref.~\cite{kouwenhoven12}. 
As for the zero-bias peak, the gap-closing feature in the differential conductance will also be significantly enhanced by disorder in the barrier region.
This is shown in Fig.~\ref{fig:gaps}b where both the zero-bias peak and the peaks associated with the topological gap for $B<B_c$ are much more pronounced than in Fig.~\ref{fig:gaps}a. Indeed, the topological gap originates from the same subband as the zero-bias peak and its visibility is thus enhanced by the same mechanism.
Given that the predictions of the multiband model (\ref{BdG}) are consistent with the experimental data of Ref.~\cite{kouwenhoven12}, the deliberate introduction of subband mixing would be an instructive probe of Majorana bound states. 

We acknowledge useful discussions with J.\ Alicea, Y.\ Oreg, G.\ Refael, and M.\ Wimmer. We are grateful for financial support by the Deutsche Forschungsgemeinschaft through SPP 1285, by the Alexander von Humboldt foundation, and by the Studienstiftung des dt.\ Volkes.

\end{document}